\documentclass[10pt]{article}

\usepackage{amsmath}
\usepackage{amssymb}

\usepackage{graphicx}

\usepackage{cite}

\usepackage{color}

\usepackage[labelfont=bf,labelsep=period,justification=raggedright]{caption}

\makeatletter
\renewcommand{\@biblabel}[1]{\quad#1.}
\makeatother

\date{}

\usepackage{afterpage}

\newcommand{\subscript}[1]{\ensuremath{_{\textrm{#1}}}}

\begin{document}

\begin{flushleft}
{\Large
\textbf{Gateway vectors for efficient artificial gene assembly {\itshape in vitro} and expression in yeast {\itshape Saccharomyces cerevisiae}}
}
\vspace{0.5cm}
\\
Claudiu V. Giuraniuc$^{1}$, 
Murray MacPherson$^{1}$, 
Yasushi Saka$^{1,\ast}$
\\
\vspace{0.5cm}
\bf{1} School of Medical Sciences, Institute of Medical Sciences, University of Aberdeen, Aberdeen, Scotland, UK
\\
$\ast$ E-mail: y.saka@abdn.ac.uk\end{flushleft}

\section*{Abstract}

Construction of synthetic genetic networks requires the assembly of DNA fragments encoding functional biological parts in a defined order. 
Yet this may become a time-consuming procedure.
To address this technical bottleneck, we have created a series of Gateway shuttle vectors and an integration vector, which facilitate the assembly of artificial genes and their expression in the budding yeast \textit{Saccharomyces cerevisiae}. 
Our method enables the rapid construction of an artificial gene from a promoter and an open reading frame (ORF) cassette by one-step recombination reaction \emph{in vitro}. 
Furthermore, the plasmid thus created can readily be introduced into yeast cells to test the assembled gene's functionality. 
As flexible regulatory components of a synthetic genetic network, we also created new versions of the tetracycline-regulated transactivators tTA and rtTA by fusing them to the auxin-inducible degron (AID).
Using our gene assembly approach, we made yeast expression vectors of these engineered transactivators,  AIDtTA and AIDrtTA and then tested their functions in yeast.
We showed that these factors can be regulated by doxycycline and degraded rapidly after addition of auxin to the medium. 
Taken together, the method for combinatorial gene assembly described here is versatile and would be a valuable tool for yeast synthetic biology.


\section*{Introduction}

Recent progress in synthetic biology has made it possible to engineer gene regulatory networks with predictable behaviors \cite{Hasty:2002uq}.
Construction of \emph{de novo} gene networks requires efficient assembly of DNA sequences including promoters and ORFs encoding proteins with defined functions. 
Yet such DNA assemblies are often time-consuming, typically involving \textit{ad hoc} procedures such as restriction digestions and DNA ligations.
An alternative method for gene assembly is the multisite Gateway recombination cloning \cite{Sasaki:2004fk}.
This system enables rapid and highly efficient construction of an expression vector containing multiple DNA fragments \textit{in vitro} from the so-called Entry clones and a Destination vector. 
A recent study applied this cloning technology successfully to the yeast one-hybrid system \cite{Deplancke:2004uq}. 
In this system a library of DNA bait sequences and ORFs encoding the reporters (His3 and $\beta$-galactosidase) were assembled in a one-step Gateway recombination reaction. 
The cloned plasmids were then integrated at the \textit{HIS3} and \textit{URA3} loci for screening \cite{Deplancke:2004uq, Reece-Hoyes:2011kx, Reece-Hoyes:2011fk}. 
However, the Gateway vectors for the one-hybrid system were not designed as versatile plasmid vectors for more general use to build artificial genes and introduce them into yeast. 
Such vectors should facilitate the rapid characterization of new promoter-ORF combinations prior to the synthesis of an artificial genetic network in yeast.

In an effort to aid gene network construction in budding yeast \emph{Saccharomyces cerevisiae}, we have adopted an approach similar to the one-hybrid system and created Gateway Destination vectors. 
These new Destination vectors are based upon the pRS series of yeast centromere plasmids \cite{Sikorski:1989fk} and an integration vector \cite{Sadowski:2007fk}.
In addition, we have constructed a set of Entry clones harboring promoters and ORFs. 
An artificial gene from a promoter and an ORF Entry clones can be assembled in a Destination vector by one-step recombination reaction \emph{in vitro}. 

To increase the flexibility of regulation of synthetic genetic networks in yeast, we engineered two transcriptional transactivators, AIDtTA and AIDrtTA. 
Each of these are a fusion of well-characterized functional protein domains, the auxin-inducible degron (AID) \cite{Nishimura2009} and the tetracycline-regulated transactivator (tTA and rtTA) \cite{Gossen:1995fk}.
tTA activity is downregulated in the presence of tetracycline antibiotics such as doxycycline, while rtTA is activated by the antibiotic.
In the AID system, a plant hormone auxin binds to Tir1, an F-box protein, which in turn induces rapid degradation of the target protein fused to AID \cite{Nishimura2009}.
Using our Gateway vectors and the recombination cloning method, we constructed expression vectors of AIDtTA and AIDrtTA and tested their functions in yeast.
We also constructed a destabilized version of Venus (enhanced YFP) ORF, yEVenus-Cln2\subscript{PEST}-NLS, to be expressed as a fluorescent reporter protein.
Taken together, these resources provide a technological platform to assemble and test a new promoter-ORF combination efficiently in yeast.


\section*{Results}

\subsection*{Construction of yeast Gateway vectors for one-step gene assembly}

To construct Gateway Destination vectors for assembling a promoter and an ORF, we cloned the DNA sequences for site-specific recombination and terminators for gene expression into the pRS series of yeast centromere plasmids \cite{Sikorski:1989fk} and an integration vector \cite{Sadowski:2007fk}.
Table \ref{tab:label2} is the list of Destination vectors created in this study (see Materials and Methods for details). 
We also made a collection of promoter and ORF Entry clones by Gateway BP recombination reactions (Table \ref{tab:label3}). 
The constructed Destination vectors (Fig. \ref{dest}) can be recombined with a promoter Entry clone and an open reading frame (ORF) Entry clone in a one-step Gateway LR reaction \textit{in vitro} (Fig. \ref{LR}). 
These promoters and ORFs were cloned in the Gateway Donor vectors in a specific direction so that assembled genes in expression vectors are always in the same orientation relative to the recombination sites, in the order of attB2-[promoter]-attB5-[ORF]-attB1-[\textit{TEF} or \textit{CYC1} terminator]. 
The attB5 sequence (5'-CAACTTTTGTATACAAAGTTG-3') has no noticeable side effect on the gene expression from vectors so far created by this method.
A recent study has also applied the Gateway system successfully for a yeast one-hybrid system, in which a single attB1 recombination sequence is created between a promoter and an ORF as a by-product of the recombination event \cite{Deplancke:2004uq, Reece-Hoyes:2011kx, Reece-Hoyes:2011fk}.
The attB5 'scar' left behind by LR reactions may therefore be unlikely to interfere with the assembled gene's expression.

One of the Destination vector we have created (pDEST375) allows integration of the artificial gene made by a Gateway recombination reaction at the \textit{MET15} locus (Fig. \ref{integration}).
This construct is based on the integration vector pIS375, which allows a single-copy integration of the gene at the \textit{MET15} locus\cite{Sadowski:2007fk}. 
The integrated constructs are highly stable because a duplicated copy of \textit{MET15} and the flanking plasmid sequences including the \textit{URA3} marker gene are removed by homologous recombination \cite{Sadowski:2007fk} (Fig. \ref{integration}).
It enables a recycling of \textit{URA3} marker gene for subsequent gene integrations at other loci in the genome. 
pDEST375 was used for the integration of TEF-AIDrtTA gene (see below and Materials and Methods for details).
For the detailed description of this class of integration/disintegration vectors, see Sadowski \emph{et al.} \cite{Sadowski:2007fk}.

We have performed a number of Gateway LR reactions using these plasmids with various combinations of promoters and ORFs to create Yeast/E. coli shuttle vectors. 
The vectors thus created for this study are listed in Table \ref{tab:label4}. 
An example of the expression vector \textit{CUP1}-yEVenus-Cln2\subscript{PEST}-NLS construct (pCM25) is shown in Fig. \ref{cup1}A.
The \textit{CUP1} promoter can be induced by copper \cite{Mateus:2000uq, Janke:2004fk}.
yEVenus-Cln2\subscript{PEST}-NLS is a fusion of the enhanced yellow fluorescent protein Venus (yEVenus), the Cln2\subscript{PEST} (which destabilizes the protein fused to it \cite{Mateus:2000uq}) and the SV40 nuclear localization signal (NLS).
The expression of the fluorescence reporter yEVenus-Cln2\subscript{PEST}-NLS was observed after one hour following the addition of the inducer copper nitrate (Fig. \ref{cup1}A, middle panels). 
Similar level of fluorescence was also detected at six hours (Fig. \ref{cup1}A, bottom panels). 
When the culture media was replaced with the one without copper after induction for two hours, the fluorescence disappeared by six hours after the initial induction (data not shown).
We compared the fluorescence produced by yEVenus-Cln2\subscript{PEST}-NLS and yEVenus-NLS when expressed by the constitutive promoter ADH1 (Fig. \ref{cup1}B, C).
The fluorescence by ADH1-yEVenus-Cln2\subscript{PEST}-NLS (Fig. \ref{cup1}B) was much weaker than that by ADH1-yEVenus-NLS (Fig. \ref{cup1}C), indicating that Cln2\subscript{PEST} indeed destabilizes yEVenus.

All but a few expression vectors have no non-specific effect on the cell growth: we have found that the yeast cells with \textit{TEF-tTA} (pCG112), and \textit{TEF-AIDtTA} (pCG106; described below) grow poorly with an unknown reason (data not shown).
It has been reported that an autoregulatory construct of tTA with a \textit{TetO\subscript{7}} promoter causes cell growth retardation when introduced in the hamster glioblastoma cell line HJC-15 \cite{Gallia:1998fk}. 
The mechanism of this delay of cell proliferation remains unclear.

\subsection*{Construction and characterization of tunable transactivators AIDtTA and AIDrtTA}

To fuse an AID to the N-terminus of tTA and rtTA, we have used a method that combines a Gateway BP reaction \emph{in vitro} and a homologous recombination \emph{in vivo} in {\emph E. coli} \cite{Suzuki:2005vn} (See Materials and Methods for details).
To verify the auxin-dependent degradation of AIDtTA and AIDrtTA, the yeast strain (YS114) with an integrated copy of \emph{O. sativa TIR1} gene {(\textit{ADH1-OsTIR1}) was transformed with a vector harboring \textit{ADH1-AIDtTA} gene (pDHM19). 
The transformed cells were grown in liquid medium and treated with 1-naphthaleneacetic acid (NAA, a synthetic analog of auxin) or left untreated. 
Cells were then harvested 0.5, 3 and 4 hours after the addition of NAA, and the expression of AIDtTA was examined by Western blotting using a monoclonal anti-TetR antibody (Fig. \ref{western}A and B; note that different anti-TetR and anti-Myc antibodies were used in panel A and B ). 
As shown in Fig. \ref{western}A, AIDtTA was efficiently degraded by 3 hours following the addition of NAA.
This degradation depended on the expression of OsTIR1 (tagged with 9xMyc epitope).
The AID was also required for the degradation as the expression of untagged tTA was not affected in YS114 strain in the presence of NAA (Fig. \ref{western}B).
Likewise, AIDrtTA was degraded in NAA-dependent manner in YS114 strain (with OsTIR1), but not in YS129 (without OsTIR1) (Fig. \ref{western}C).
When AIDrtTA was  expressed by the strong TEF promoter in YS114 strain, a fraction of AIDrtTA protein remained intact even 4 hours after the addition of NAA to the medium (Fig. \ref{western}D).

Next, we verified the activity of AIDtTA as a transcriptional activator.
YS129 yeast strain was transformed with the reporter plasmid construct pCG87 (\textit{TetO\subscript{7}-mCherry-NLS}; mCherry fluorescent protein reporter construct) together with pCG84 (\textit{ADH1-tTA}) or pDHM19 (\textit{ADH1-AIDtTA}).
As shown in Fig. \ref{AIDtTA}, tTA and AIDtTA are active in the absence of doxycycline (-DOX).
Because tTA or AIDtTA activity is supposed to be suppressed in the presence of DOX, mCherry reporter fluorescence should be reduced after treating cells with DOX.
After 7 hours following the addition of DOX to the medium, however, significant fluorescence still remained (Fig. \ref{AIDtTA}B and C, +DOX).
This is presumably because the half-life of mCherry is relatively long and it may take longer to see the reduction of the fluorescence after addition of DOX, which is consistent with the results of flow cytometry (see below).

We asked whether any rapid reduction of fluorescence after the suppression of AIDtTA activity by DOX and NAA can be achieved using yEVenus-Cln2\subscript{PEST} as a reporter. 
For this, we used a diploid strain (JW003) with a copy of \textit{ADH1-OsTIR1} and a reporter \textit{TetO\subscript{7}-yEVenus-Cln2\subscript{PEST}} genes integrated in the genome.
JW003 was transformed with pDHM19 (\textit{ADH1-AIDtTA}).
Note that both \textit{AIDtTA} and \textit{OsTIR1} was expressed by constitutively active \textit{ADH1} promoter.
The yeast cells were then cultured in the presence or absence of NAA or DOX alone, or both together.
Fig. \ref{AIDtTAvenus}A shows a schematic diagram of the gene regulation by AIDtTA in this strain. 
Without the inducers DOX or NAA, Venus (YFP) fluorescence was clearly visible in the nucleus (Fig. \ref{AIDtTAvenus}B).
Cells without the plasmid (pDHM19; \textit{ADH1-AIDtTA}) showed no fluorescence (data not shown).
After 1 hour of treatment with DOX, NAA or both, no obvious decrease of fluorescence was observed (Fig. \ref{AIDtTAvenus}C, upper left panels).
The addition of DOX took effect in 2 hours, with much reduced fluorescence in the presence of the inducer either alone or together with NAA (Fig. \ref{AIDtTAvenus}C, upper right panels).
It took 3 hours for NAA alone to  suppress effectively the expression of the reporter yEVenus-Cln2\subscript{PEST} (Fig. \ref{AIDtTAvenus}C, lower left panels).
By 4 hours after the addition of the inducers, fluorescence was greatly reduced in all cases (Fig. \ref{AIDtTAvenus}C, lower right panels).
The cells treated with both DOX and NAA showed more pronounced reduction of fluorescence than those with either DOX or NAA alone.
These results indicate that AIDtTA activity can be controlled efficiently by DOX and NAA. 
DOX took effect faster than NAA perhaps partly because DOX directly inhibits the reporter gene induction by AIDtTA while NAA acts indirectly by promoting the degradation of AIDtTA.
We tested two concentrations of NAA at 0.5mM and 2mM: increasing the concentration of NAA did not accelerate the reduction of the fluorescence (the results with 2mM NAA is shown in Fig. \ref{AIDtTAvenus}).
We also performed a similar experiment using YS114 strain (with an integrated \textit{ADH1-OsTIR1}) transformed with two plasmids pDHM19 (\textit{ADH1-AIDtTA}) and pCG103 (\textit{TetO\subscript{7}-yEVenus-Cln2\subscript{PEST}}): similar profile of reduction in fluorescence was observed after cells were treated with DOX and NAA, although the level of fluorescence was more heterogeneous across the cell population in all conditions tested than that in JW003 strain (data not shown).

The same set of yeast strains used in Fig. \ref{AIDtTA}, i.e., YS129 or YS114 with two vectors pCG87 (mCherry reporter plasmid) and pCG84 (\textit{ADH1-tTA}) or pDHM19 (\textit{ADH1-AIDtTA}) were subjected to flow cytometry to examine the gene expression control by AIDtTA (Fig. \ref{facs}A-C).
Cells were treated with either doxycycline (+DOX; broken green lines) or NAA alone (+NAA; cyan), or both (+DOX \& NAA; orange) for 7 hours, or left untreated (no inducers; broken magenta).
Consistent with the observation by fluorescence microscopy (Fig. \ref{AIDtTA}), tTA and AIDtTA were equally active in the absence of doxycycline and NAA (broken magenta lines in Fig. \ref{facs}A-C), and induced the expression of the mCherry reporter fluorescent protein.
After addition of DOX (broken} green lines), the fluorescence of the reporter decreased as expected for both  strains expressing tTA and AIDtTA.
A large fraction of cells, however, have the fluorescence well above the basal level of the reporter plasmid ($<10^2$ arbitrary fluorescence units, data not shown).
In the presence of NAA (cyan lines), the mCherry expression decreased only in the cells with both AIDtTA and OsTIR1 expressed (Fig. \ref{facs}C).
These results are consistent with the NAA- and OsTIR1-dependent degradation of AIDtTA detected by Western blots (Fig. \ref{western}A and B).

We also tested the doxycycline-dependent gene expression by AIDrtTA and its degradation in OsTIR1- and NAA-dependent manner (Fig. \ref{AIDrtTA}). 
The YS114 (with OsTIR1) and YS129 (without OsTIR1) strains transformed with the plasmids pCG87 (mCherry reporter construct) and pCG107 (\textit{TEF-AIDrtTA}) (Fig. \ref{AIDrtTA}A) were cultured in the presence or absence of DOX or NAA for 5 hours, and observed by fluorescence microscopy (Fig. \ref{AIDrtTA}B). 
Expression of the fluorescent reporter mCherry was detected in response to +DOX in both YS114 and YS129 strains (Fig. \ref{AIDrtTA}B, middle panels).
In contrast, when DOX was added to the cell together with NAA (+DOX\&NAA), the reporter expression was inhibited in YS114 strain (Fig. \ref{AIDrtTA}B, right bottom panels).
The expression of mCherry was not affected in the absence of OsTIR1 after addition of NAA (Fig. \ref{AIDrtTA}B, left bottom panels).
DOX-dependent expression of reporter fluorescent proteins was also observed using the strain with an integrated copy of TEF-AIDrtTA (MM017; Fig. \ref{AIDrtTA}C).

Those yeast strains used for fluorescence microscopy were also examined by flow cytometry (Fig. \ref{facs}D-F).
The results confirmed the DOX-dependent activation (green broken line) by AIDrtTA (Fig \ref{facs}E and F).
The data also demonstrated the NAA-dependent reduction (orange line) of fluorescent reporter proteins in cells expressing AIDrtTA and OsTIR1 (Fig \ref{facs}F).
In Fig \ref{facs}D-F, the data for cells treated with NAA alone are not shown, which have similar distribution of fluorescence as those of untreated ones (No inducers; broken magenta lines). 
YS129 (the strain without OsTIR1) with \textit{TEF-AIDrtTA} showed around five fold decrease in fluorescence when DOX was added together with NAA (Fig \ref{facs}E, green broken and orange lines); the reason for this reduction in fluorescence remains unknown.
We also found that when rtTA expression was induced by the strong promoter \textit{TEF}, it activated the reporter gene in the absence of DOX (Fig \ref{facs}D, magenta broken line).
Such spurious DOX-independent activation of the \textit{TetO} promoter by rtTA was not observed when its expression was induced by moderate \textit{ADH1} promoter (data not shown).

\section*{Discussion}

In this study, we have created a set of versatile Gateway Destination vectors and a collection of promoter and ORF Entry clones to facilitate gene network construction. 
The multisite Gateway technology enables the efficient assembly of a promoter and an ORF in a one-step recombination reaction \textit{in vitro}. 
An expression vector generated by this method can readily be introduced into yeast for functional assays of the assembled gene. 
A similar combinatorial approach was demonstrated to be an efficient method for producing diverse phenotypes of synthetic networks in \emph{E. coli} \cite{Guet:2002ys}. 
The Destination vector pDEST375 described above allows a stable single-copy integration of an artificial gene at \textit{MET15} locus, adding a flexible option for engineering artificial gene networks in yeast. 
We plan to construct similar Gateway vectors for integration at the other chromosome locus utilized by Sadowski \textit{et al.} \cite{Sadowski:2007fk}, such as \textit{ADE8}, \textit{FCY1} and \textit{LYS2}.

The collection of Entry clones for our gene assembly strategy may be expanded to include a number of other biological parts which has already been characterized and available for synthetic biology applications. 
For instance, DNA fragments encoding a subset of BioBrick standard biological parts (http://partsregistry.org/) \cite{Canton:2008fk} could be converted into Entry clones by PCR amplification using the prefix and suffix sequences that flank every BioBrick part, followed by Gateway BP recombination reactions.
Although the Entry clones presented here are for two-fragment recombinations, the Destination vectors can be used for recombination reactions of three or more DNA fragments. This would be useful for the assembly of a gene with a complex promoter.

In an effort to increase the flexibility of synthetic gene networks in yeast, we have also constructed two new tunable transcriptional transactivators, AIDtTA and AIDrtTA. 
We showed that when \emph{O. sativa TIR1} gene is co-expressed, they were rapidly degraded in response to auxin.
We also demonstrated that the expression of the reporter gene by AIDtTA and AIDrtTA were controlled efficiently by NAA and/or DOX added to the medium (Fig. \ref{AIDtTA}-\ref{AIDrtTA}).
We constructed a destabilized version of Venus, yEVenus-Cln2\subscript{PEST}, and used it to monitor the AIDtTA's activity (Fig. \ref{AIDtTAvenus}). 
Because Venus has faster, improved maturation properties of the chromophore \cite{Nagai:2002uq}, yEVenus-Cln2\subscript{PEST} may be better-suited than the original version, GFP-Cln2\subscript{PEST} \cite{Mateus:2000uq}, as a reporter of the dynamics of synthetic gene networks.
A unique advantage of AIDtTA or AIDrtTA may be that the 'memory' of gene activation can be erased rapidly by an auxin-regulated protein degradation.
The rapid degradation of these transactivators are however dependent on the fine balance of their expression level and that of Tir1 protein: when AIDrtTA was expressed by the strong TEF promoter, its degradation was retarded (Fig. \ref{western}D).
The AID system therefore requires fine-tuning of the expression level of its components to achieve the desired functionality.
For example, the expression level of OsTIR1 in yeast may be increased by adopting a codon-optimized version of the ORF \cite{Kubota:2013fk}, thereby improving the efficiency of the degradation of AID-tagged target proteins.

An additional control may also be integrated into synthetic networks using these engineered transactivators by making the \textit{TIR1} gene expression inducible.
AIDtTA and AIDrtTA should also be functional in other eukaryotic systems in principle, as AID system and the tetracycline-regulated transactivators were demonstrated to work in higher eukaryotic cells \cite{Nishimura2009,Gossen:1995fk}.
Because of the dominance of the auxin-induced degradation over the doxycycline-regulated gene induction, AIDtTA and AIDrtTA may function as Boolean logic devices in response to doxycycline and auxin, corresponding to NOR and N-IMPLY gate, respectively.
Recent work has also exploited post-transcriptional negative regulations to perform logic operations NOT, AND, NAND and N-IMPLY, using RNA binding proteins that inhibit the translation of transcripts harboring specific RNA motifs \cite{Auslander:2012uq}.
AIDtTA and AIDrtTA may be combined together with other biological logic operation devices in eukaryotes \cite{Regot:2011kx,Rodrigo:2012vn,Auslander:2012uq,Lohmueller:2012ly}. 

The budding yeast \emph{Saccharomyces cerevisiae} is an attractive system for synthetic biology because of its vast resources and tools for biotechnology applications.  
Moreover, \emph{S. cerevisiae} has been exploited as a platform for synthesizing a complete genome \cite{Gibson:2008uq,Tagwerker:2012fk} and chromosome arms \cite{Dymond:2011kx}.
These studies have presented a prospect of engineering a large-scale artificial gene network in yeast. 
Although the number of biological parts customized for yeast synthetic biology is still limited, it is rapidly growing in recent years \cite{Ellis:2009ys,Blount:2012vn,Khalil:2012zr}.
The multisite Gateway recombination method for assembling artificial genes may be a useful addition to the toolbox for synthetic biology in yeast and also a valid approach in other organisms.

\section*{Materials and Methods}

\subsection*{Yeast manipulations, strains and media}

Genetic manipulations and transformations of yeast cells were performed as described \cite{cslmanual}. 
After transformation, cells were plated on synthetic complete medium (SC) lacking appropriate amino acids \cite{cslmanual}. 
Transformed cells with plasmids were cultured in the selective media until the culture reaches early to mid-log phase of growth ($OD\subscript{600} = 0.1 \sim 1.0$) for experiments.
The yeast strains used in this study are FY1679 ($MAT\textbf{a}/\alpha~GAL2/GAL2~HIS3/his3\Delta 200~LEU2/leu2\Delta 1~TRP1/trp1\Delta63~ura3-52/ura3-52$; EUROSCARF Acc. No. 10000D), YS114 ($MAT\alpha~ GAL2~ his3\Delta 200~ trp1\Delta 63~ ura3-52::ADH1-OsTIR1-9Myc, URA3$; this study), MM017 ($MAT\textbf{a}~ his3\Delta1~ leu2\Delta 0~ trp1\Delta 63~ ura3\Delta 0~ met15::TEF-AIDrtTA$; this study), YS008 ($MAT\alpha~  his3\Delta 1~  leu2\Delta 0~ trp1\Delta 63~ ura3\Delta 0~ met15::ADH1-OsTIR1-9Myc$; this study),  YS017 ($MAT\textbf{a}~ his3\Delta 1~  leu2\Delta 0~ trp1\Delta 63~ ura3\Delta 0~ lys2::TetO\subscript{7}-CYC1TATA-yEVenus-Cln2\subscript{PEST}-NLS$; this study), YS081 ($MAT\alpha~  his3\Delta 1~  leu2\Delta 0~ trp1\Delta 63~ ura3\Delta 0$; this study), JW003 ($MAT\textbf{a}/\alpha~ his3\Delta 1/his3\Delta 1~  leu2\Delta 0/leu2\Delta 0~ trp1\Delta 63/trp1\Delta 63~ ura3\Delta 0/ura3\Delta 0~ met15::ADH1-OsTIR1-9Myc/MET15~ LYS2/ lys2::TetO\subscript{7}-CYC1TATA-yEVenus-Cln2\subscript{PEST}-NLS$; this study) and YS129 ($MAT\textbf{a}~ his3\Delta 1~  leu2\Delta 0~ trp1\Delta 63~ ura3\Delta 0$; this study), which are all congenic with S288C. 
To integrate the \emph{Oryza sativa} \textit{TIR1} gene, pNHK53 (an integration vector of \textit{ADH1-OsTIR1-9Myc}) \cite{Nishimura2009} was linearized by StuI and transformed into FY1679. 
To derive YS114, the diploid strain with an integrated pNHK53 was sporulated. Sporulations were performed as follows: Freshly streaked cells on YPD plate were patch-streaked on GNA (5\% glucose, 3\% Difco nutrient broth (BD Bioscience; cat. no. 234000), 1\% yeast extract, 2\% bacto-agar) and incubated overnight at $30^{\circ}C$. 
The cells on GNA were patch-streaked again on GNA and  incubated overnight at $30^{\circ}C$. 
A small quantity of cells (equivalent to a medium-sized colony on a plate) was then suspended in 2.5ml of sporulation medium (1\% pottasium acetate, 0.5\% zinc acetate) and incubated at $25^{\circ}C$ with vigorous shaking for up to a week.
The genetic selection of YS008, YS017 and MM017 strains were performed according to Sadowski \emph{et al.} \cite{Sadowski:2007fk}, which describes the detailed methods of gene integration and disintegration.
To create MM017, pMM6 (Table \ref{tab:label4}) was linearized with MluNI (an isoschizomer of MscI) and transformed into YS129 for integration.
To derive YS008, pCM33 (integration vector of \textit{ADH1-OsTIR1-9Myc}; see below for its construction) was linearized by MluNI and transformed into YS129 for integration.
To make YS017, pCM36 (\textit{TetO\subscript{7}-CYC1TATA-yEVenus-Cln2\subscript{PEST}-NLS}; see below for its construction) was linearized by NruI and transformed into YS081 strain.
JW003 strain was made by mating YS008 and YS017. JW003 transformed with the plasmid pDHM19 (ADH1-AIDtTA) was used for the experiment shown in Fig. \ref{AIDtTAvenus}.

To test the regulation of AIDtTA and AIDrtTA by doxycycline and auxin (Fig. \ref{western} to \ref{AIDrtTA}), YS114, MM017 and YS129 strains were transformed with pCG103 (\textit{TetO\subscript{7}-CYC1TATA-yEVenus-Cln2\subscript{PEST}-NLS}), pCG87 (\textit{TetO\subscript{7}-CYC1TATA-mCherry-NLS}), pCG84 (\textit{ADH1-tTA}), pDHM19 (\textit{ADH1-AIDtTA}), pCG85 (\textit{ADH1-rtTA}), pDHM20 (\textit{ADH1-AIDrtTA}), pCG113 (\textit{TEF-rtTA}), and pCG107 (\textit{TEF-AIDrtTA}) as indicated in each figure.
Doxycycline (DOX) was added to cell cultures at the final concentration of $5~\mu g/ml$.
For the induction of protein degradation using AID system, 1-Naphthaleneacetic acid (NAA; pottasium salt, Sigma cat. no. N1145) was added to cell cultures  to 0.5mM, except the experiment shown in Fig. \ref{AIDtTAvenus} in which 2mM was used.
We observed little inhibition of cell growth up to 2mM of NAA. 
Phosphate-citrate buffer (64.2mM Na\subscript{2}HPO\subscript{4} and 17.9mM citric acid, pH6.0) was added to SC medium to prevent the acidification of the culture \cite{Murakami:2011fk} when using NAA.
To induce{ \itshape CUP1} promoter, copper nitrate (Cu(NO\subscript{3})\subscript{2}, Sigma cat. no. 223395) was added to the medium to 0.25mM.
DOX, NAA, or copper nitrate were added to cell cultures at OD\subscript{600} = 0.2.

\subsection*{Molecular biology techniques}

Standard molecular biology techniques were used for DNA manipulations. DNA fragments were purified using a PCR purification kit or gel purification kit (Qiagen). 
DNA ligation was performed using a rapid DNA ligation kit (Roche Applied Science). 
For PCR reactions, Pfu polymerase (Promega) or Velocity DNA polymerase (Bioline) were used. 
PCR primers used in this study are listed in Table \ref{tab:label1}. 
Plasmid clones created with PCR reactions were verified by DNA sequencing. 
The Gateway recombination reactions were performed using Multisite Gateway Pro Kits (Life Technologies). 
The donor vectors (pDONR221P1-P5r, pDONR221P5-P2; Life Technologies), and destination vectors were prepared using ccdB Survival $T1^R$ Chemically Competent \textit{E. coli} (Life Technologies).\\
\indent{\bf Construction of Destination vectors.}
The destination vectors containing a yeast centromere were built upon the pRS vector series (pRS413, pRS414, pRS415 and pRS416) \cite{Sikorski:1989fk}.
\textit{A. gossypii TEF} terminator was amplified by PCR from pAG25 \cite{Goldstein:1999uq}, using the standard T7 sequencing primer and TEFt-F primer. 
The amplified DNA was cut with NotI and SacI and cloned into the pRS vectors to create pRS413TEFt, pRS414TEFt, pRS415TEFt and pRS416TEFt.
\textit{S. cerevisiae CYC1} terminator was amplified from pCM183 \cite{Gari:1997kx} by PCR with CYC1F and CYC1R primers, cut with NotI and SacI and cloned into the pRS413 vector digested with the same enzymes to create pRS413CYC1T.
A cassette containing two recombination sites, attR1 and attR2, flanking the \textit{ccdB} and $Cm^R$ (chloramphenicol resistance) genes was amplified by PCR from pDEST17 (Invitrogen), using primers attR1-F and attR2-R. 
The PCR product was digested by SpeI and AvrII and cloned into pRS413CYC1T to yield pDEST413CYC1T.
The attR1-attR2 cassette with the \textit{ccdB} and $Cm^R$ genes was cut out from pDEST413CYC1T by XhoI and XbaI and cloned into pRS413TEFt, pRS415TEFt and pRS416TEFt, which yielded pDEST413TEFt, pDEST415TEFt and pDEST416TEFt, respectively.
To create pDEST414TEFt, 2kb DNA fragment containing the cassette and the \textit{TEF} terminator was cut out from pDEST413TEFt by XhoI and SacI and ligated to pRS414 digested with the same enzymes.
Similarly, the XhoI-SacI DNA fragment from pDEST413CYC1T was cloned into pRS414 to make pDEST414CYC1t.
Although these pDEST plasmids were functional as destination vectors, T7 terminator downstream of \textit{ccdB} gene was found to be required for efficient selection of expression clones.
Therefore, DNA fragments harbouring the T7 terminator were cut out from pDEST17 and cloned into these vectors, to create pDEST413TEFt7, pDEST414TEFt7, pDEST415TEFt7, pDEST416TEFt7, pDEST413CYC1t7 and pDEST414CYC1t7 (Fig.\ref{tab:label2}). 
pDEST375 was made by isolating the KpnI/SacI DNA fragment containing the Gateway attR1-attR2 cassette from pDEST416TEFt7 and cloning it into pIS375 \cite{Sadowski:2007fk} using the same restriction sites. \\
\indent{\bf Construction of promoter Entry clones.}
\textit{S. cerevisiae CUP1} promoter was amplified by PCR from pYM-N1 \cite{Janke:2004fk} using the oligonucleotide primers B2-CUP1-F and B5-CUP1-R. 
\textit{S. pombe ADH1} promoter was amplified from pCM245 \cite{Belli:1998uq} using B2-ADH-F and B5-ADH-R primers. 
\textit{S. cerevisiae TEF} promoter was amplified from pYM-N18 \cite{Janke:2004fk} using the primers B2-TET-F and B5-TET-R. 
A DNA fragment containing \textit{TetO\subscript{2}} operator with the \textit{TATA} region of \textit{S. cerevisiae CYC1} gene (\textit{CYC1TATA}) was amplified from pCM171 \cite{Gari:1997kx} using the primers B2TetO-1 and B5TetO-1. 
Likewise, \textit{TetO\subscript{7}} with \textit{CYC1TATA} was amplified from pCM159 \cite{Gari:1997kx} using the same pair of primers. 
These PCR products were purified and subjected to BP recombination reaction with pDONR221P5-P2 \emph{in vitro} to create the plasmids listed in Table \ref{tab:label2}.\\
\indent{\bf Construction of ORF Entry clones.}
yEGFP ORF was amplified from pYM25 \cite{Janke:2004fk} using the primers B5r-yEGFP-F2 and B1-yEGFP-R2.
yEGFP-Cln2\subscript{PEST} ORF was obtained by PCR from pSVA13 \cite{Mateus:2000uq} using B5r-yEGFP-F2 and B1-Cln2PEST.
yEVenus ORF was amplified by PCR from pKT90 \cite{Sheff:2004fk} using the primers B5r-yEGFP-F2 and B1-yVenus-R2.
mCherry ORF were amplified from pCS-memb-mCherry (mCherry fused to a myristoylation motif; a kind gift from J.C. Smith, originally constructed by S. Megason and S. Fraser, Caltech), using B5r-mCherry-F1 and B1-mCherry-R1 primers.
A glycine linker and a SV40 nuclear localization signal (NLS; amino acid sequence PGGGGPKKKRKVD) was added to the C-terminus of yEVenus ORF by PCR using the primers B5r-yEGFP-F2 and B1-yEGFPNLS-R and pKT90 as a template.
Likewise, mCherry-NLS was constructed by PCR using the primer pair B5r-mCherry-F1 and B1-mCherryNLS-R, and pCG40 as a template.
The plasmid for yEVenus-Cln2\subscript{PEST} (pCG98) was created by inserting a Cln2\subscript{PEST} XmaI fragment into pYS61. 
The Cln2\subscript{PEST} sequence was amplified by PCR using primers XmaI-PESTF and XmaI-PESTR, and pSVA13 as a template.
tTA and rtTA ORFs were amplified from pCM171 \cite{Gari:1997kx} and pCM251 \cite{Belli:1998uq} respectively, using the primers B5r-tTA and B1-tTA.
To create pCG72, OsTIR1-9Myc was amplified by PCR using primers B5r-TIR1 and B1-TIR1, and pNHK36 \cite{Nishimura2009} as a template.
The amplified ORFs were cloned into pDONR221P1-P5r by Gateway BP reaction to create the Entry clones in Table \ref{tab:label3}. 
Auxin inducible degron (AID) was fused to tTA and rtTA using a method that combines a Gateway BP reaction \emph{in vitro} and a homologous recombination \emph{in vivo} in \textit{E. coli} \cite{Suzuki:2005vn}. For this, two PCR reactions were performed.
First, a DNA fragment encoding AID was amplified from pNHK12 \cite{Nishimura2009} using the oligonucleotides AIDtTA-R and B5r-AIDGFPNLS (PCR product 1). 
Second, tTA and rtTA ORF were amplified using the primers AIDtTA-F and B1-tTA (PCR product 2).
PCR product 1 and 2 have an overlapping sequence that mediates a homologous recombination \emph{in vivo}.
These PCR products together with pDONR221P1-P5r were subjected to a BP reaction, followed by a transformation into \textit{E. coli} ($DH5\alpha$ strain), to create pYS57 (Entry clone of AIDrtTA) and pYS58 (AIDtTA).\\
\indent{\bf Other plasmids.}
The integration vector pCM33 was made as follows: 
pCG81 (Table \ref{tab:label4}) was digested by SphI/BspQI and blunt-ended by T4 DNA polymerase (New England Biolabs). 
The fragment containing \textit{ADH1-OsTIR1-9Myc} was purified and ligated to pIS375 \cite{Sadowski:2007fk}, which was linearized by BamHI and blunt-ended by Klenow fragment (Promega). 
pCM36 was created by inserting PvuII fragment of pCM20 (Table \ref{tab:label4}) into pIS385 \cite{Sadowski:2007fk}, which was cut by BamHI and blunt-ended by Klenow fragment.
 pNHK12, pNHK36 and pNHK53 were obtained from the National Bio-Resource Project (NBRP) of MEXT, Japan.
The following plasmids were obtained from EUROSCARF (EUROpean Saccharomyces Cerevisiae ARchive for Functional Analysis): pAG25, pCM159, pCM171, pCM183, pCM245, pCM251, pIS375, pKT90, pYM25, pYM-N1, pYM-N18. 

\subsection*{Microscopy and flow cytometry}

Yeast cells were cultured in liquid SC medium lacking appropriate amino acids and mounted on a glass slide for fluorescence microscopy. 
Images were acquired using a Zeiss Axioplan2 microscope with an oil-immersion objective lens (100x) equipped with a CCD camera (Hamamatsu ORCA-ER) and Openlab software (Perkin Elmer).
For flow cytometry, cells were harvested by centrifugation, or by vacuum filtration on nitrocellulose membranes ($1.2 \mu m$, Millipore, cat. no. RAWP02500), re-suspended in ice-cold 1xPBS and sonicated briefly before subjected to the analysis using LSRFortessa (BD Biosciences). 
Fluorescence of 50,000 cells for each sample were measured by  flow cytometry. 
Based on the FFS/SSC values, the data were gated to remove cell debris and aggregates. 
The raw data for the remaining cells, between 86\% and 98\% of the original sample, were exported to MATLAB (MATLAB, Natick, Massachusetts: The MathWorks Inc., 2012) and binned in log scale ($0~to~10^5, 100~bins$). 
The scale of each bin was normalized by the sum of all cells in the respective sample and plotted (Fig. \ref{facs}). 
No smoothing has been applied.

\subsection*{Western blotting}

 Protein extraction from yeast cells was conducted as described \cite{Kushnirov:2000fk}. 
 Standard techniques were used for electrophoresis (SDS-PAGE) and Western blotting. 
 Western blotting was performed using Amersham ECL Plus Western Blotting detection reagent (cat no. RPN2132). 
 The primary antibodies used were anti-myc monoclonal (1:1000, Millipore, 9E10, *2 in Fig. \ref{western}A) or rabbit polyclonal antibody (1:5000, Abcam, cat no. ab9106, *4 in Fig. \ref{western}B-D) and TetR monoclonal antibody (1:1000, Clontech, cat no. 631108, *1 in Fig. \ref{western}A; 1:1000, MoBiTec, TET02, *3 in Fig. \ref{western}B-D); the secondary antibodies were anti-mouse IgG HRP-linked antibody (1:1000, Cell Signaling \#7076) and anti-rabbit IgG HRP-linked antibody (1:1000, DAKO, cat no. P0217). 
 Equal loading of protein extracts in SDS-PAGE was verified by Ponceau S staining of blotted membranes. 

\section*{Acknowledgments}

We thank Philip Hieter and Simon Avery for providing plasmids; Hakima Abes, Droth\'{e}e Vicogne, Yvonne Turnbull, Diane Massie, Claire Macgregor, Linda Duncan, Barry Lewis and Kimberley Sim for technical assistance. 
We also thank Marta Morawska, Simon Jackson, Richard Newton and Jareth Wolfe for preliminary results. 
The flow cytometry was completed with assistance from the university of Aberdeen flow cytometry core facility. 
We thank CNRS (Centre National de Recherche Scientifique) for supporting the early stage of this project.
This work was funded by SULSA (Scottish Universities Life Sciences Alliance; http://www.sulsa.ac.uk/).

\bibliography{plos}


\section*{}
\afterpage{
\section*{Figures}

\begin{figure}[!h]
\begin{center}
\includegraphics[width=8.3cm]{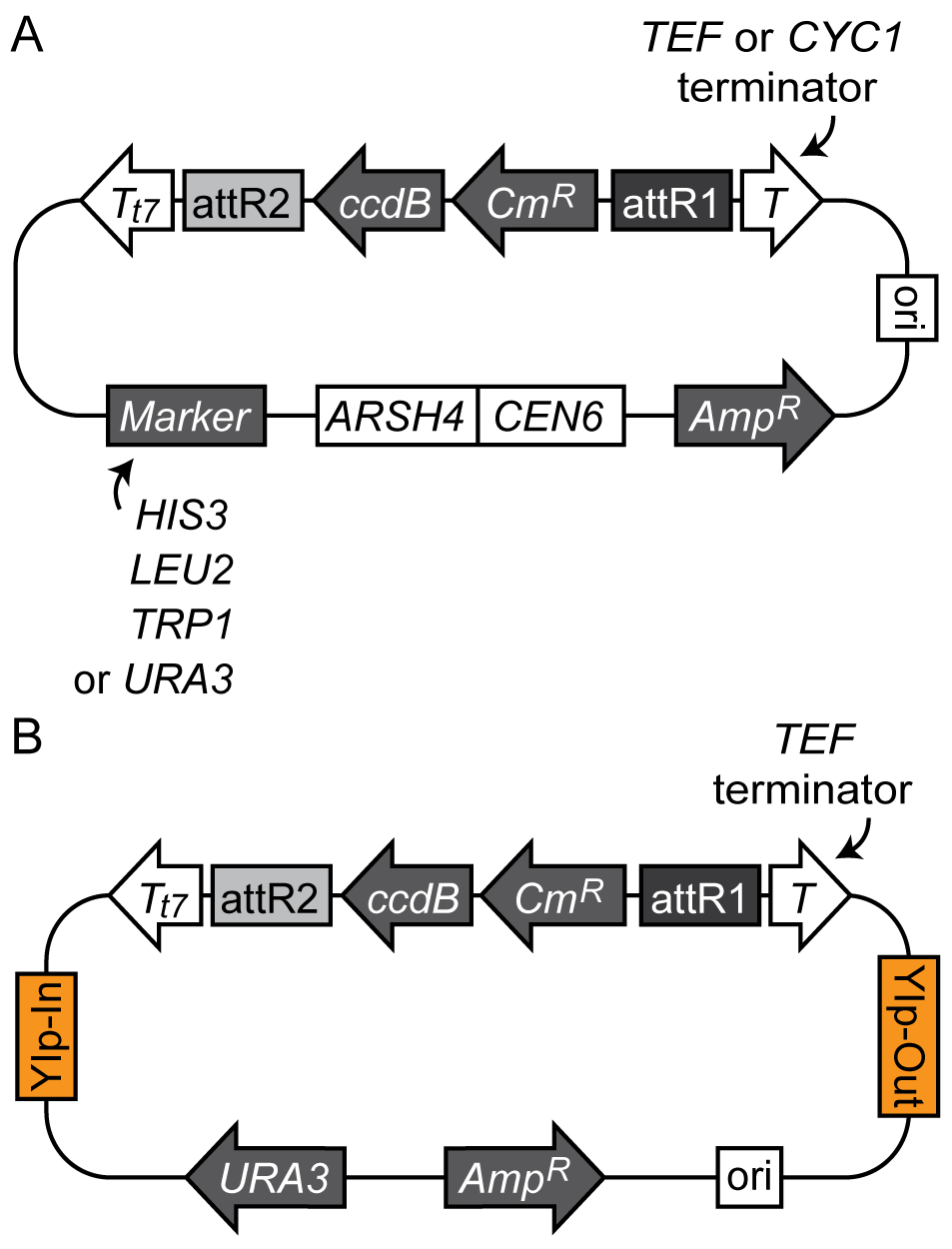}
\end{center}
\caption{
{\bf Gateway Destination vectors constructed in this study.} 
(A) Yeast centromere Destination vectors. Each vector contains \textit{ARSH4}, \textit{CEN6}, and a marker gene for maintenance in yeast. (B) Yeast integration Destination vector (pDEST375). All vectors have an origin of replication in \textit{E.coli} (ori), attR1 and attR2 recombination sequences for Gateway LR reaction, ampicillin resistance ($Amp^R$), chloramphenicol resistance ($Cm^R$) and \textit{ccdB} selection marker genes for cloning. attR1 and attR2 are flanked by \textit{TEF} or \textit{CYC1} terminator (T) and T7 terminator (T\subscript{t7}) sequences as indicated.
}
\label{dest}
\end{figure}

\begin{figure}[t]
\begin{center}
\includegraphics[width=12.35cm]{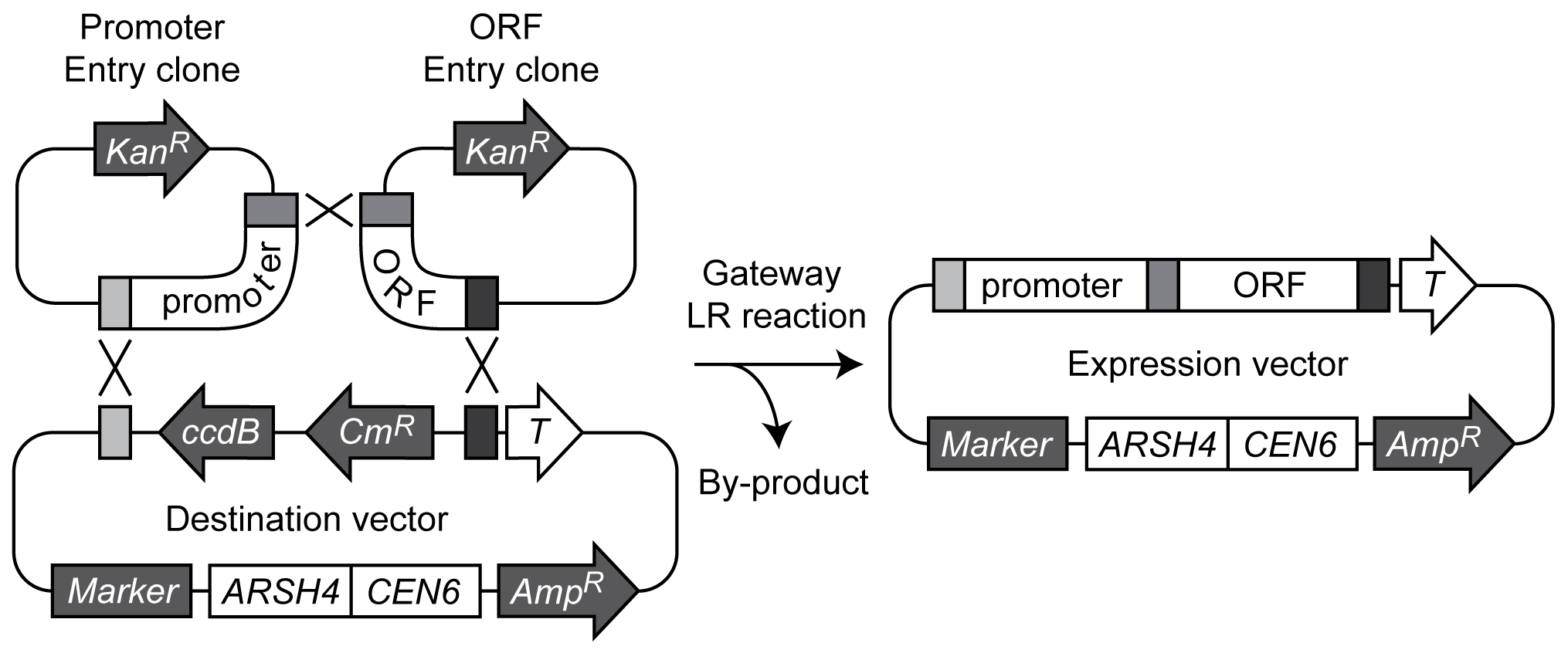}
\end{center}
\caption{
 {\bf Gene assembly by Gateway LR recombination reaction.}  Yeast expression vectors can be created by one-step recombination reaction using a promoter entry clone, an ORF entry clone and a Destination vector. In the LR reaction, $Cm^R$ and \textit{ccdB} selection marker genes flanked by attR1 and attR2 are replaced by the assembled gene (promoter-ORF).
 }
\label{LR}
\end{figure}

\newpage
      
\begin{figure}[!h]
\begin{center}
\includegraphics[width=12.35cm]{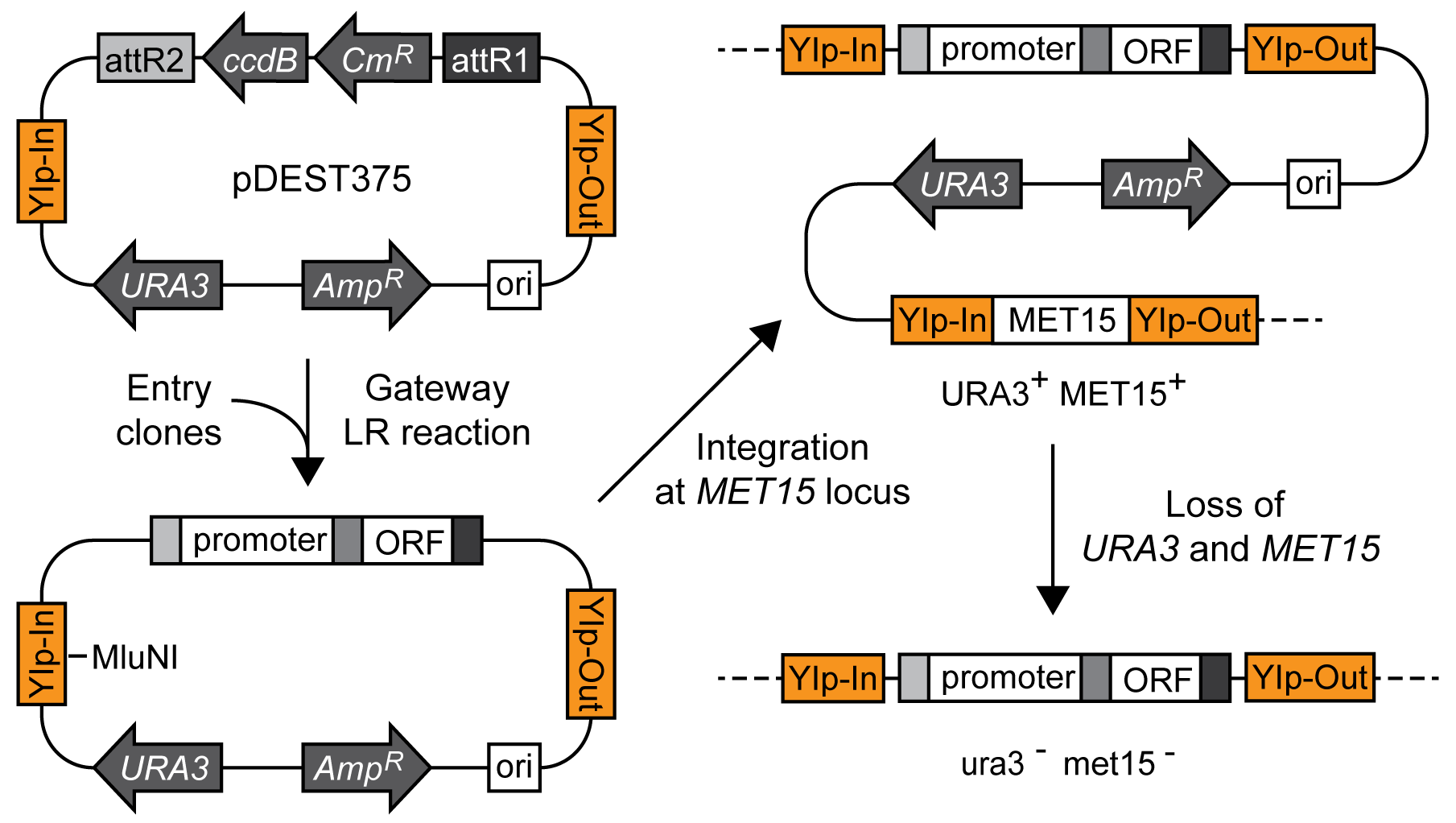}
\end{center}
\caption{
{\bf Gateway vector pDEST375 for chromosomal integration.} pDEST375 harbors YIp-In and YIp-Out sequences derived from the flanking region of MET15 gene. An integration vector made by an LR reaction is first linearized by MluNI (MscI), which is transformed into yeast cells. It produces a single-copy integration of the gene of interest at the \textit{MET15} locus. At this stage, the genotype is $URA3^{+}MET15^{+}$. The chromosomal \textit{MET15} gene and the flanking plasmid sequences including the \textit{URA3} marker gene are removed by homologous recombination between the duplicated YIp-Out sequences. After this recombination event \emph{in vivo}, the desired genotype of the strain with the stable integrated gene (\textit{promoter-ORF}) is $ura3^{-}met15^{-}$. The other possible genotype without the integrated gene after the recombination is $ura3^{-}MET15^{+}$ (strains with this genotype will be discarded). It enables a recycling of \textit{URA3} marker gene for subsequent gene integrations at other loci in the genome.
}
\label{integration}
\end{figure}

\newpage

\begin{figure}[c]
\begin{center}
\includegraphics[width=15.35cm]{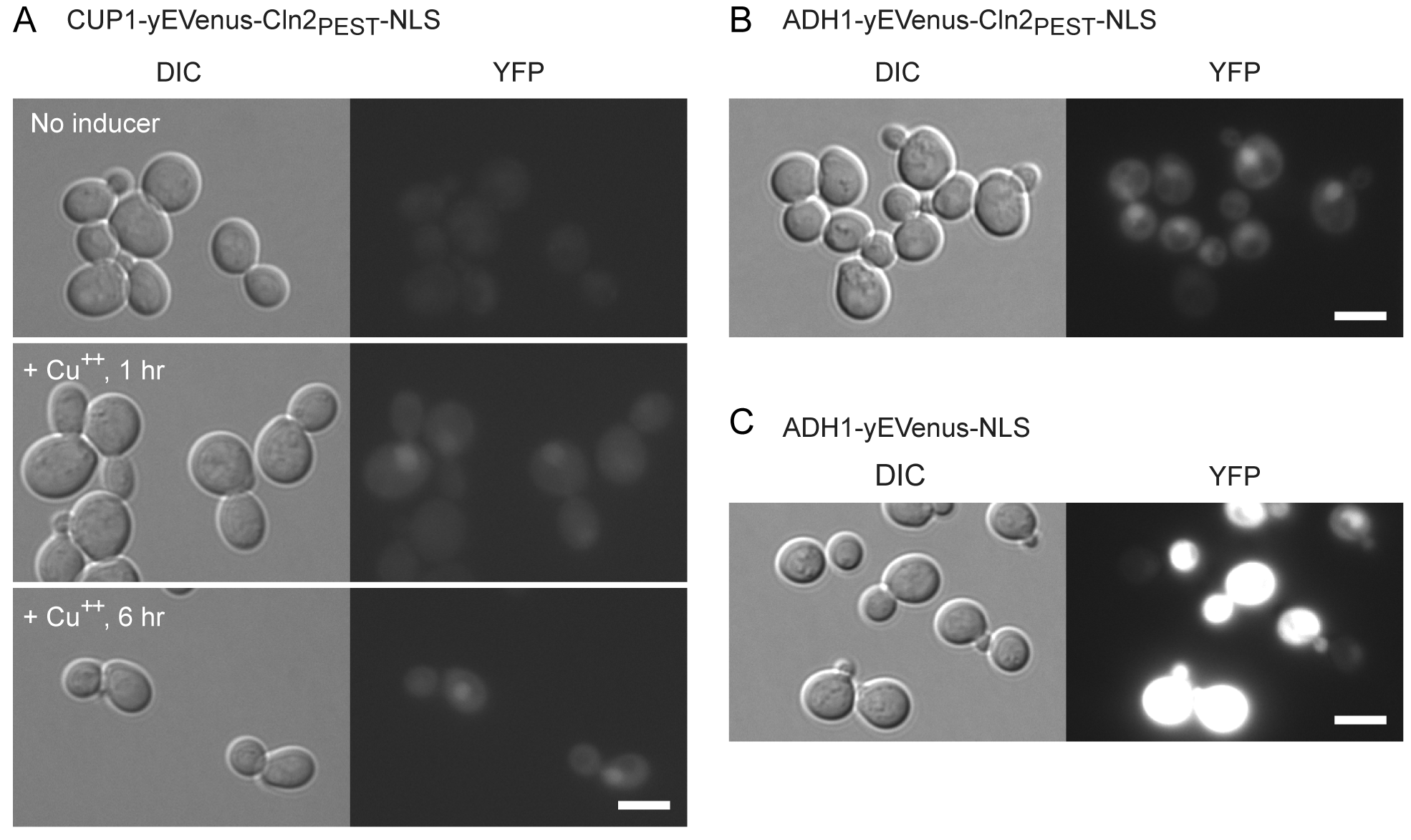}
\end{center}
\caption{
{\bf Fluorescent protein gene induction by CUP1 and ADH1 promoter.}  (A) Yeast cells (YS129 strain) transformed with pCM25 (\textit{CUP1}-yEVenus-Cln2\subscript{PEST}-NLS) were cultured in the presence or absence of copper nitrate for 1 hr (middle panels) and 6 hrs (bottom panels). Cells with no copper nitrate are also shown in the upper panel. (B) Yeast cells (YS129) with pDHM57 (\textit{ADH1}-yEVenus-Cln2\subscript{PEST}-NLS). (C) Yeast cells (YS129) with pRN1 (\textit{ADH1}-yEVenus-NLS). Note that the fluorescence images in (B) and (C) were captured with the same exposure time. DIC: differential interference contrast; YFP: YFP filter channel for the detection of Venus fluorescence. $Scale~bar = 5~\mu m$.
}
\label{cup1}
\end{figure}

\newpage

\begin{figure}[!ht]
\begin{center}
\includegraphics[width=15.35cm]{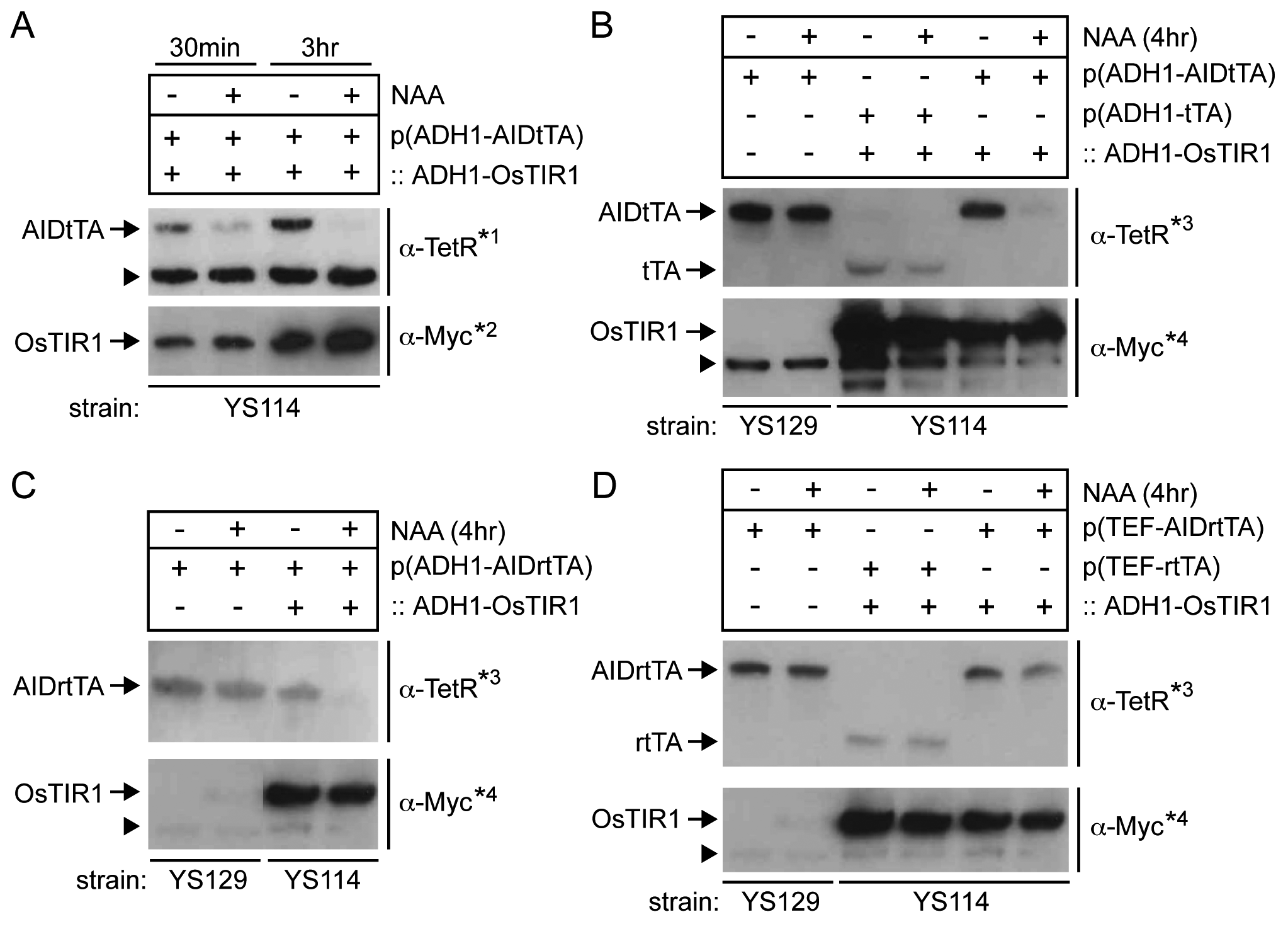}
\end{center}
\caption{
{\bf Auxin-induced degradation of AIDtTA and AIDrtTA detected by Western blotting.} 
OsTIR1 is tagged with 9xMyc epitope. Cells were treated with NAA or left untreated for the periods as indicated. 
(A) The yeast strain with an integrated copy of \textit{OsTIR1} gene (::ADH1-OsTIR1; YS114 strain) and \textit{ADH1-AIDtTA} plasmid. 
(B) Control strain YS129 with \textit{ADH1-AIDtTA}, YS114 with \textit{ADH1-tTA}, and YS114 with \textit{ADH1-AIDtTA} plasmid. 
(C) YS129 and YS114 with \textit{ADH1-AIDrtTA} plasmid.
(D) YS129 with \textit{TEF-AIDrtTA}, YS114 with \textit{TEF-rtTA} and YS114 with \textit{TEF-AIDrtTA} plasmid. 
Cells were harvested and subjected to Western blotting using anti-TetR ($\alpha-TetR$, for tTA, AIDtTA, rtTA and AIDrtTA) and anti-Myc ($\alpha-Myc$, for OsTIR1) antibodies. 
Note that different primary antibodies were used in Fig. \ref{western}A (indicated by *1 and *2) and B-D (*3 and *4); see Materials and Methods. 
Arrowheads indicate non-specific proteins recognized by the primary antibodies, which act as loading controls.
}
\label{western}
\end{figure}
   
\newpage

\begin{figure}[c]
\begin{center}
\includegraphics[width=15.35cm]{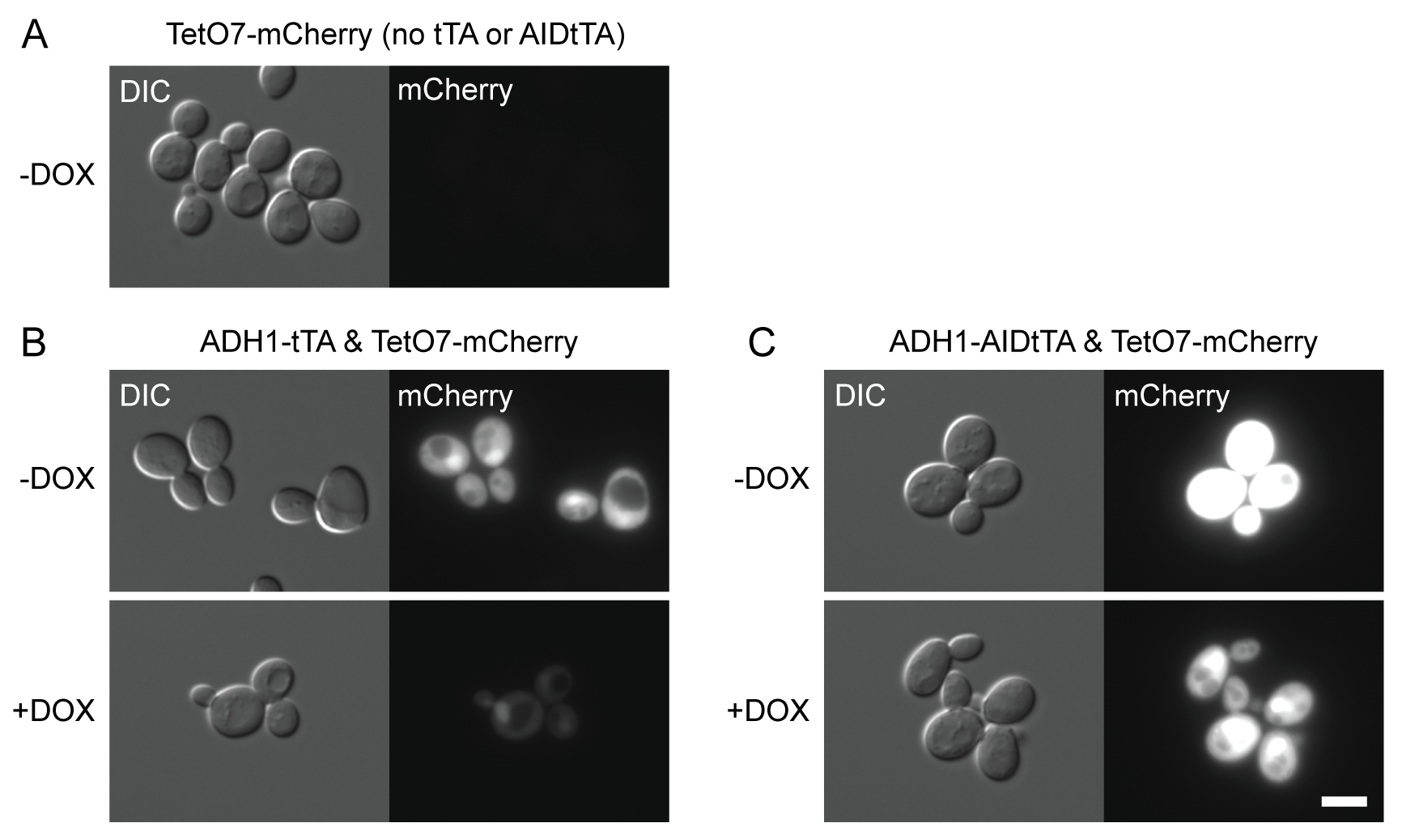}
\end{center}
\caption{
 {\bf Fluorescent reporter protein expression induced by AIDtTA.}  YS129 strain was transformed with the plasmid constructs indicated above each panel. The activity of tTA and AIDtTA was detected by mCherry fluorescent reporter protein expression. (A) TetO\subscript{7}-mCherry-NLS (reporter plasmid) only, without doxycycline (-DOX). (B) ADH1-tTA and TetO\subscript{7}-mCherry-NLS. (C) ADH1-AIDtTA and TetO\subscript{7}-mCherry-NLS. Cells were observed 7 hrs after addition of DOX. DIC: differential interference contrast; mCherry: mCherry fluorescence was detected using TRITC filter. $Scale~bar = 5~\mu m$.
 }
\label{AIDtTA}
\end{figure}

\newpage

\begin{figure}[!ht]
\begin{center}
\includegraphics[width=12.35cm]{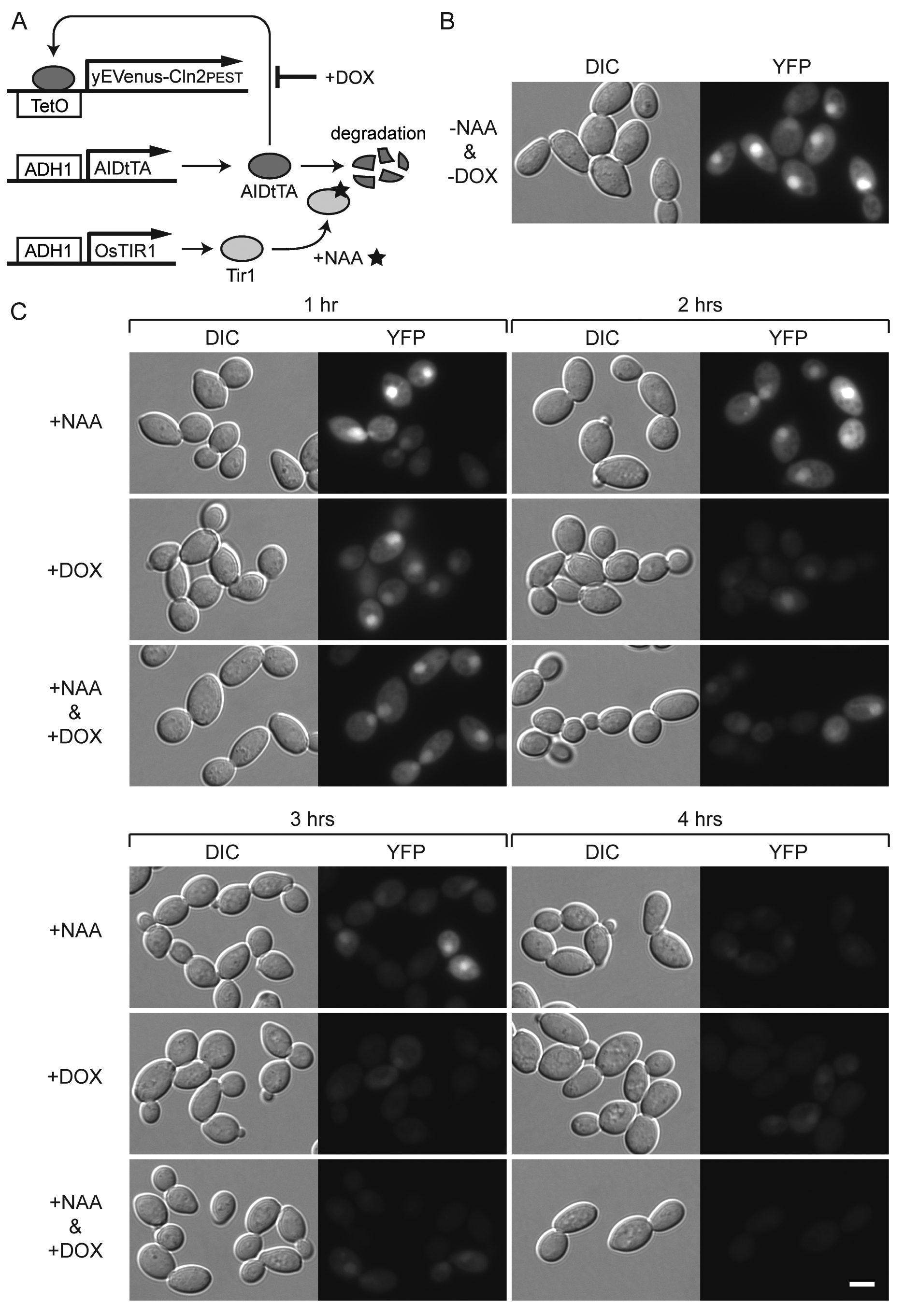}
\end{center}
\caption{
{\bf Suppression of AIDtTA activity by doxycycline and NAA.}
The yeast strain with an integrated copy of \textit{ADH1-OsTIR1} and a reporter \textit{TetO\subscript{7}-yEVenusCln2\subscript{PEST}-NLS} gene was transformed with the plasmid harboring \textit{ADH1-AIDtTA}. The fluorescent reporter protein expression was examined in the presence or absence of DOX and NAA as indicated. (A) Schematic diagram of the gene network of the strain. (B) Expression of yEVenusCln2\subscript{PEST} in the absence of DOX and NAA. (C) Expression of yEVenusCln2\subscript{PEST} after treatment with DOX or NAA alone or both for 1, 2, 3 and 4 hours. DIC: differential interference contrast; YFP: YFP filter channel. $Scale~bar = 5~\mu m$.
}
\label{AIDtTAvenus}
\end{figure}


\begin{figure}[!ht]
\begin{center}
\includegraphics[width=15.35cm]{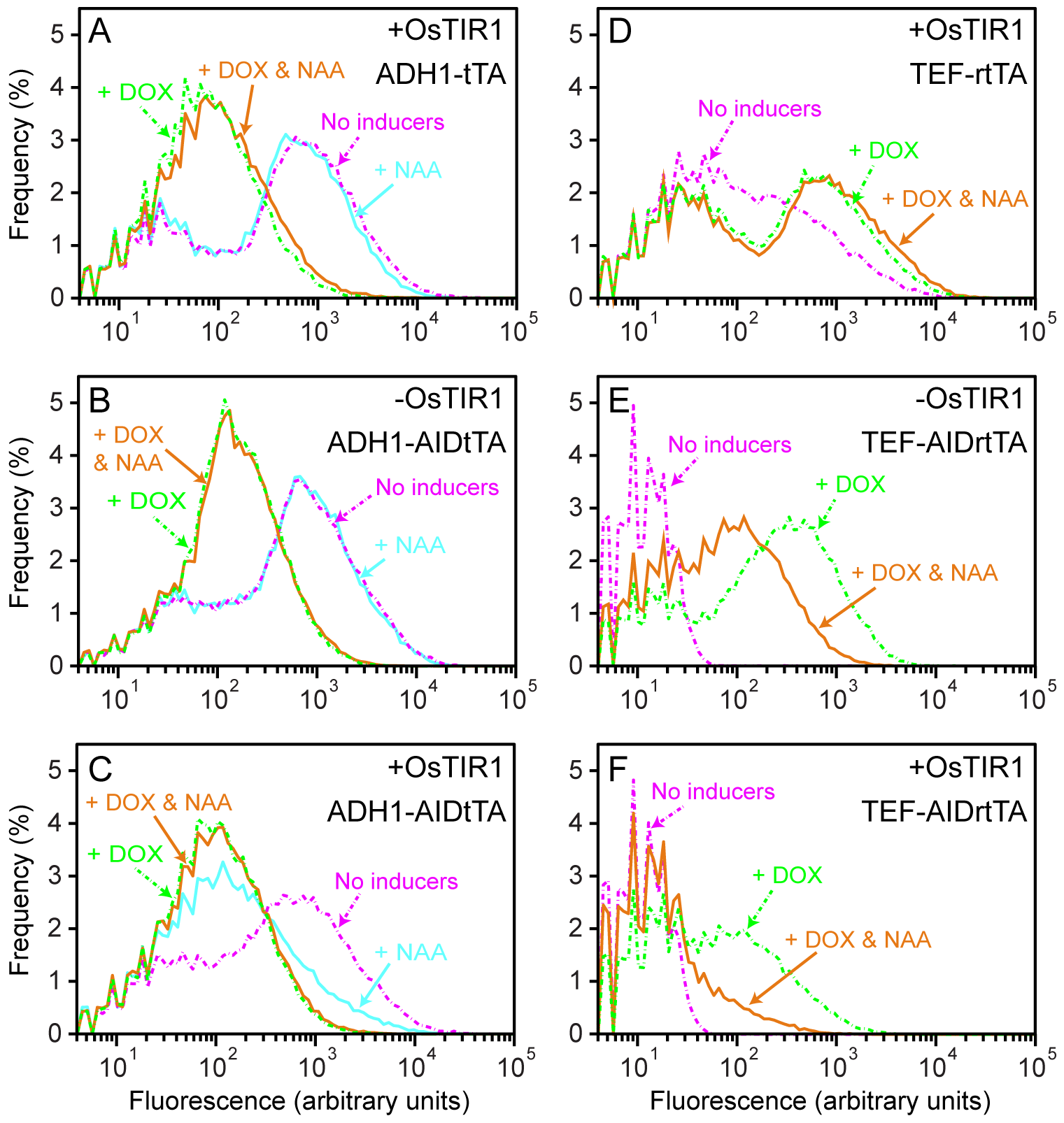}
\end{center}
\caption{
{\bf Gene expression control by AIDtTA and AIDrtTA: flow cytometry analysis.}  (A-F) The yeast strains YS129 (-OsTIR1) and YS114 (+OsTIR1) were transformed with a reporter plasmid construct (TetO\subscript{7}-mCherry-NLS) together with the plasmid harboring the gene as labeled on each panels.  These transformed yeast cells were treated with either doxycycline (+DOX; broken green lines) or NAA alone (+NAA; cyan), or both (+DOX \& NAA; orange) for 7 hours, or left untreated (no inducers; broken magenta), and analyzed by flow cytometry. 
Fluorescence of 50,000 cells for each sample were measured.
The data were gated to remove those of cell debris and aggregates, binned in log scale ($0~to~10^5, 100~bins$) and plotted.
The plots are offset at fluorescence = 4 arbitrary units.
}
\label{facs}
\end{figure}


\begin{figure}[!ht]
\begin{center}
\includegraphics[width=15.35cm]{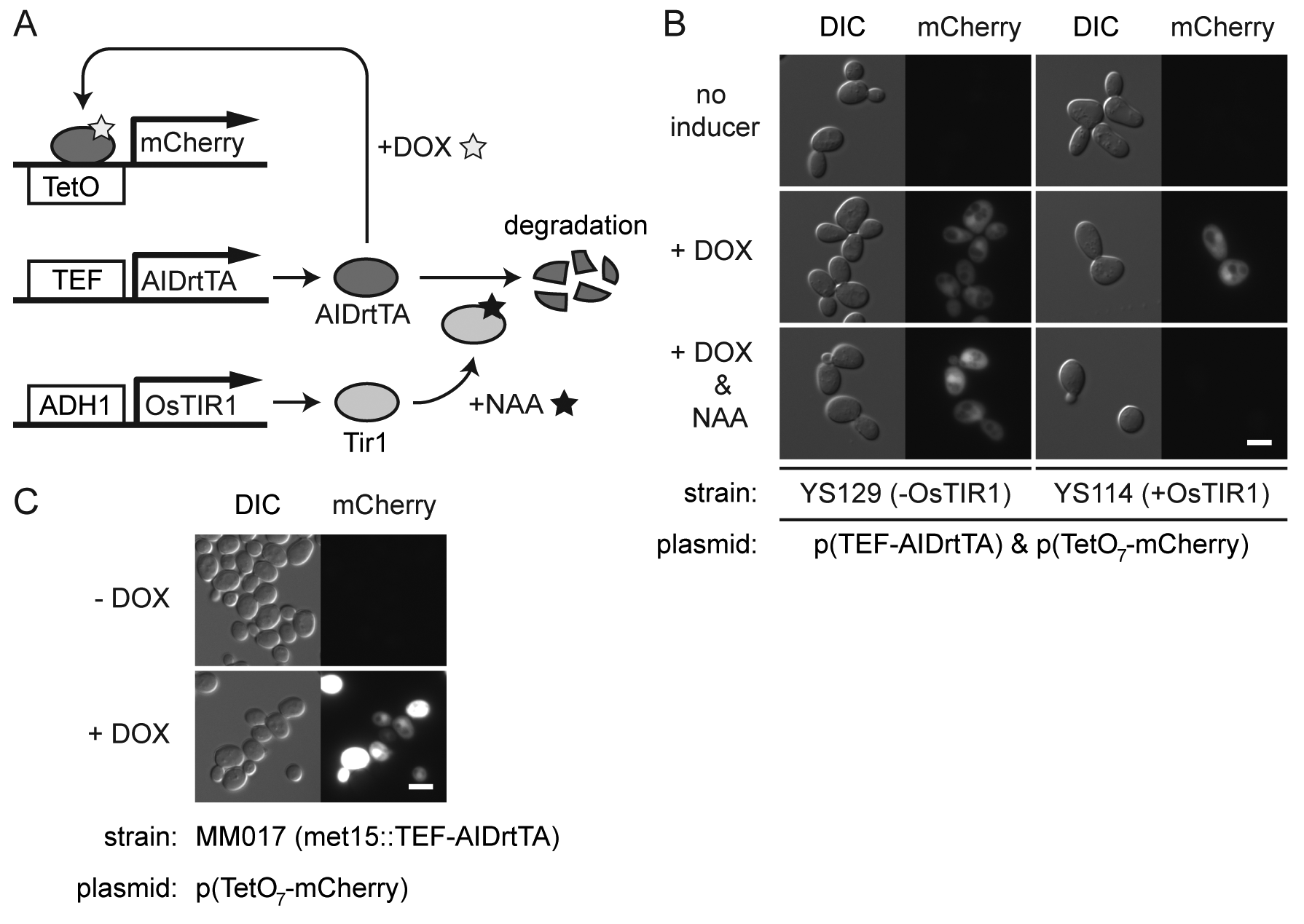}
\end{center}
\caption{
{\bf Gene expression control by AIDrtTA.}  (A) Schematic representation of the experimental system. AIDrtTA activity was detected by mCherry fluorescent reporter protein expression. (B) Yeast strains as indicated were cultured in the presence of doxycycline (+DOX), auxin (+NAA), both doxycycline and auxin (+DOX \& NAA), or in the absence (no inducer). Cells were observed 6 hrs after addition of the inducers. 
   (C) Yeast strain with an integrated copy of \textit{TEF-AIDrtTA} (MM017) was cultured in the presence or absence of doxycycline for 5 hours. DIC: differential interference contrast; mCherry: mCherry fluorescence detected with TRITC filter. $Scale~bar = 5~\mu m$.
}
\label{AIDrtTA}
\end{figure}

\clearpage
}

\newpage
\afterpage{
\section*{Tables}

\begin{table}[!h]
\caption{
\bf{Destination vectors created in this study}
}
\begin{tabular}{ l c c}
\hline
{\bf  Plasmid} &{\bf Marker}&{\bf GenBank Acc. No.}\\
\hline
pDEST413TEFt7 & \textit{HIS3} & JX901379\\
pDEST414TEFt7 & \textit{TRP1} & JX901380\\
pDEST415TEFt7 & \textit{LEU2} & JX901381\\
pDEST416TEFt7 & \textit{URA3} & JX901382\\
pDEST413CYC1t7 & \textit{HIS3} & JX901383\\
pDEST414CYC1t7 & \textit{TRP1} & JX901384\\
pDEST375 & $URA3, {MET15}^{*1}$ & KC614689\\
\hline

\end{tabular}

\begin{flushleft}
\textsuperscript{*1}\,5' and 3' flanking sequences of \textit{MET15}.
\end{flushleft}

\label{tab:label2}
\end{table}

\begin{table}[!h]
\caption{
\bf{Entry clones created in this study}
}
\begin{tabular}{ l c }
\hline
{\bf  Plasmid} &{\bf Promoter}\\
\hline
pYS1 & \textit{S. cerevisiae CUP1}\\
pYS2 & \textit{S. pombe ADH1}\\
pYS3 & \textit{S. cerevisiae TEF}\\
pYS6 & \textit{TetO\subscript{7}-CYC1TATA}\\
pYS7 & \textit{TetO\subscript{2}-CYC1TATA}\\
\hline
{\bf  Plasmid} & {\bf ORF}\\
\hline
pCG32 & yEGFP\\
pDHM7 & yEGFP-Cln2\subscript{PEST}\\
pCG55 & yEVenus\\
pYS61 & yEVenus-NLS$^{*1}$\\
pCG98 & yEVenus-Cln2\subscript{PEST}-NLS\\
pCG40 & mCherry\\
pYS60 & mCherry-NLS\\
pYS19 & tTA\\
pYS20 & rtTA\\
pYS58 & AIDtTA\\
pYS57 & AIDrtTA\\
pCG72 & OsTIR1-9Myc\\
\hline
\end{tabular}
\begin{flushleft}
Promoter Entry clones were created with pDONR221P5-P2 and ORF Entry clones with pDONR221P1-P5r.
$^{*1}$ Significant cytoplasmic fluorescence was observed when overexpressed, for example, by \textit{TEF} promoter.
\end{flushleft}
\label{tab:label3}
\end{table}

\newpage
\begin{table}[t]
\caption{
\bf{Expression plasmid vectors constructed by Gateway recombination method in this study}
}
\begin{tabular}{ l c c c c}
\hline
{\bf  Plasmid} &{\bf Promoter}&{\bf ORF}&{\bf Marker} & {\bf Destination vector}\\
\hline
pCG52 &\textit{S. cerevisiaeTEF}& mCherry &\textit{LEU2} & pDEST415TEFt7\\
pCG109 &\textit{S. pombe ADH1}& mCherry-NLS &\textit{TRP1} & pDEST414TEFt7\\
pCG57 &\textit{S. cerevisiae TEF}&yEVenus&\textit{LEU2} & pDEST415TEFt7\\
pRN1 &\textit{S. pombe ADH1}&yEVenus-NLS&\textit{LEU2} & pDEST415TEFt7\\
pDHM57 &\textit{S. pombe ADH1}&yEVenus-Cln2\subscript{PEST}-NLS&\textit{LEU2} & pDEST415TEFt7\\
pCM25 & \textit{S. cerevisiae CUP1}& yEVenus-Cln2\subscript{PEST}-NLS &\textit{LEU2} & pDEST415TEFt7\\
pCG87 & \textit{TetO\subscript{7}-CYC1TATA}&mCherry-NLS&\textit{TRP1} & pDEST414TEFt7\\
pCG103 & \textit{TetO\subscript{7}-CYC1TATA}&yEVenus-Cln2\subscript{PEST}-NLS&\textit{TRP1} & pDEST414TEFt7\\
pCM20 & \textit{TetO\subscript{7}-CYC1TATA}&yEVenus-Cln2\subscript{PEST}-NLS&\textit{TRP1} & pDEST415TEFt7\\
pCG84 & \textit{S. pombe ADH1} &tTA&\textit{HIS3} & pDEST413TEFt7\\
pCG85 & \textit{S. pombe ADH1} &rtTA&\textit{HIS3} & pDEST413TEFt7\\
pDHM19 & \textit{S. pombe ADH1} &AIDtTA&\textit{HIS3}  & pDEST413TEFt7\\
pDHM20 & \textit{S. pombe ADH1} &AIDrtTA &\textit{HIS3} & pDEST413TEFt7\\
pCG112 $^{*1}$ &\textit{S. cerevisiae TEF}&tTA&\textit{HIS3} & pDEST413TEFt7\\
pCG113 &\textit{S. cerevisiae TEF}&rtTA&\textit{HIS3} & pDEST413TEFt7\\
pCG106 $^{*1}$ &\textit{S. cerevisiae TEF}&AIDtTA&\textit{HIS3} & pDEST413TEFt7\\
pCG107 &\textit{S. cerevisiae TEF}&AIDrtTA&\textit{HIS3} & pDEST413TEFt7\\
pMM6 $^{*2}$&  \textit{S. cerevisiaeTEF} & AIDrtTA & \textit{URA3, MET15} & pDEST375\\
pCG81 & \textit{S. pombe ADH1} & OsTIR1-9Myc & \textit{URA3} & pDEST416TEFt7\\
\hline
\end{tabular}
\begin{flushleft}
$^{*1}$ Yeast cells with these expression vectors showed poor growth with an unknown reason.\\
$^{*2}$ Integration vector.
\end{flushleft}
\label{tab:label4}
\end{table}

\begin{table}[t]
\caption{
\bf{PCR primers}
}
\begin{tabular}{cp{10.5cm}}
\hline
{\bf Primer} & \multicolumn{1}{c}{\bf Sequence}\\
\hline
T7 & \texttt{TAATACGACTCACTATAGGG}\\
TEFt-F & \texttt{TTGCGGCCGCTCAGTACTGACAATAAAA}\\
CYC1F &\texttt{AAGCGGCCGCATCATGTAATTAGT}\\
CYC1R &\texttt{TTGAGCTCAAATTAAAGCCTTCGAGCGT}\\
attR1-F & \texttt{ACCACTAGTACAAGTTTGTACAAAAAAGC}\\
attR2-R & \texttt{GTTCCTAGGACCACTTTGTACAAGAA}\\
B2-CUP1-F & \texttt{GGGGACCACTTTGTACAAGAAAGCTGGGTATAGTAAGCCGATCCCATTAC}\\
B5-CUP1-R & \texttt{GGGGACAACTTTGTATACAAAAGTTGCTCTGTCGTCCGGATTTATGTGATG}\\
B2-ADH-F & \texttt{GGGGACCACTTTGTACAAGAAAGCTGGGTAGCATGCCCTACAACAACTAA}\\
B5-ADH-R & \texttt{GGGGACAACTTTGTATACAAAAGTTGGCAATTCTCTTGCTTAAAGAAAAGC}\\
B2-TET-F&\texttt{GGGGACCACTTTGTACAAGAAAGCTGGGTAGAGCTCATAGCTTCAAAATG}\\
B5-TET-R&\texttt{GGGGACAACTTTGTATACAAAAGTTGAAAACTTAGATTAGATTGCTATGCT}\\
B2TetO-1& \texttt{GGGGACCACTTTGTACAAGAAAGCTGGGTAGGCAGATCAATTCCTCGATC}\\
B5TetO-1& \texttt{GGGGACAACTTTGTATACAAAAGTTGTAATTTAGTGTGTGTATTTG}\\
B5r-tTA & \texttt{GGGGACAACTTTTGTATACAAAGTTGTACAACATGTCTAGATTAGATAAAAG}\\
B1-tTA & \texttt{GGGGACAAGTTTGTACAAAAAAGCAGGCTCTACCCACCGTACTCGTCA}\\
AIDtTA-F & \texttt{GGTGCAGGCGCTGGAGCGGGTGCCATGTCTAGATTAGATAAAAGTAAA}\\
AIDtTA-R & \texttt{TTTACTTTTATCTAATCTAGACATGGCACCCGCTCCAGCGCCTGCACC}\\
B5r-AIDGFPNLS & \texttt{GGGGACAACTTTTGTATACAAAGTTGTACAAAATGATGGGCAGTGTCGAG}\\
B5r-yEGFP-F2 & \texttt{GGGGACAACTTTTGTATACAAAGTTGTACAAAATGTCTAAAGGTGAAGAATTATTC}\\
B1-yEGFP-R2 & \texttt{GGGGACAAGTTTGTACAAAAAAGCAGGCTTTATTTGTACAATTCATCCATACC}\\
B1-yVenus-R2 & \texttt{GGGGACAAGTTTGTACAAAAAAGCAGGCTTTATTTGTACAATTCATCAATACC}\\
B1-Cln2PEST & \texttt{GGGGACAAGTTTGTACAAAAAAGCAGGCTCTAGCTATATTACTTGGGTATTGC}\\
B1-yEGFPNLS-R & \texttt{GGGGACAAGTTTGTACAAAAAAGCAGGCTGGTACCTCAGTCGACTTTTCTCTTTTT-\newline CTTTGGACCACCACCCCCGGGTTTGTACAATTCATCCATACC}\\
B5r-mCherry-F1 &\texttt{GGGGACAACTTTTGTATACAAAGTTGTACAAAATGGTGAGCAAGGGCGAGGA}\\
B1-mCherry-R1 &\texttt{GGGGACAAGTTTGTACAAAAAAGCAGGCTTTACTTGTACAGCTCGTCCA}\\
B1-mCherryNLS-R & \texttt{GGGGACAAGTTTGTACAAAAAAGCAGGCTGGTACCTCAGTCGACTTTTCTCTTTTT-\newline CTTTGGACCACCACCCCCGGGCTTGTACAGCTCGTCCATGCC}\\
XmaI-PESTF & \texttt{ATACCCCGGGTCTGGTGGTACCGCATCCAACTTGAACATTTC}\\
XmaI-PESTR & \texttt{TCAACCCGGGACCACCTATTACTTGGGTATTGCCCATACC}\\
B5r-TIR1 & \texttt{GGGGACAACTTTTGTATACAAAGTTGTACAAAATGACGTACTTCCCGGAG}\\
B1-TIR1 & \texttt{GGGGACAAGTTTGTACAAAAAAGCAGGCTTTAGCTAGTGGATCCGTT}\\
\hline
\end{tabular}
\begin{flushleft}
PCR primers used in this study. Sequences are in $5^\prime~to~3^\prime$ direction.
\end{flushleft}
\label{tab:label1}
\end{table}

\clearpage
}

\end{document}